\documentclass[a4paper,12pt]{article}

\newcommand{\sect}[1]{\setcounter{equation}{0}\section{#1}}

\begin{document}
\newcommand{\beq}{\begin{equation}}
\newcommand{\eeq}{\end{equation}}
\newcommand{\beqa}{\begin{eqnarray}}
\newcommand{\eeqa}{\end{eqnarray}}
\newcommand{\sr}{\sqrt}
\newcommand{\fr}{\frac}
\newcommand{\mn}{\mu \nu}
\newcommand{\G}{\Gamma}
\newcommand{\g}{ G_{n+2}}
\topmargin 0pt
\oddsidemargin 0mm
 
\begin{titlepage}
\begin{flushright}
INJE-TP-02-04 \\
hep-th/0208086
\end{flushright}
 
\vspace{5mm}
\begin{center}
{\Large \bf Bouncing and cyclic universes in the charged AdS bulk background}
\vspace{12mm}
 
{\large
 Yun Soo Myung\footnote{email address:
 ysmyung@physics.inje.ac.kr}}
 
\vspace{8mm}
{ Relativity Research Center and School of Computer Aided Science \\ Inje University,
   Gimhae 621-749, Korea}
\end{center}
\vspace{5mm}
\centerline{{\bf{Abstract}}}
\vspace{5mm}
 
We study bouncing and  cyclic universes from  an $(n+1)$-dimensional brane 
in the $(n+2)$-dimensional charged AdS bulk background. In the moving domain 
wall (MDW) approach this picture is 
clearly realized with a specified
bulk configuration,  the 5D charged topological AdS (CTAdS$_5$) black hole with
mass $M$ and charge $Q$. The  bulk gravitational dynamics induces the 4D 
Friedmann equations with CFT-radiation and exotic stiff matters for a dynamic brane. This provides bouncing universes for $k=0,-1$ and
 cyclic universe for $k=1$, even though
it has an exotic stiff matter from the  charge $Q$.
In this work  we use the other  of the Binetruy-Deffayet-Langlos (BDL) approach 
with the bulk Maxwell field. 
In this case we are free to determine the corresponding mass $\tilde M$ and charge $\tilde Q$
because the mass term is usually included as an initial condition and the charge is given by
an unspecified solution to the Maxwell equation under the BDL metric. Here we 
obtain only bouncing universes
if one does not choose two CTAdS$_5$ black holes as the bulk spacetime.
We provide a way of  avoiding the exotic matter on the brane by introducing an appropriate local matter.
Finally we discuss an important relation between the exotic holographic  matter and 
Lorentz invariance violation.
\end{titlepage}
 
\newpage
\renewcommand{\thefootnote}{\arabic{footnote}}
\setcounter{footnote}{0}
\setcounter{page}{2}

\sect{Introduction}
 
Recently there has been much interest in cyclic (oscillating) and bouncing universes of
brane cosmology~\cite{KOST,KOSST,MP,BS}. In these studies one started 
with the moving domain wall (MDW)~\cite{Kraus,Ida,BCG,MSM}.
This is obtained by   embedding of an $(n+1)$-dimensional brane into 
the $(n+2)$-dimensional bulk background. 
In this approach
one has to choose a specified
bulk configuration, for example, the 5D charged topological AdS (CTAdS$_5$) black hole with
mass $M$ and charge $Q$~\cite{CEJM,CS}. Then  the bulk gravitational dynamics 
induces  the 4D Friedmann equations with CFT-radiation and exotic stiff matters for a 
dynamic brane~\cite{BV1,BM,Myung1,CMO,Youm}. 
This provides bouncing and cyclic universes, even though
it has an exotic stiff matter from the black hole charge $Q$~\cite{MP,Med1}.
 
 Without a charge $Q$, one finds a scenario with a CFT-radiation matter: the universe
 starts from the black hole singularity (big bang), crosses the horizon of the black hole, 
and  expands up to its maximum size. And then it contracts, crosses the horizon of the 
black hole again, and finally 
collapses into the singularity (big crunch)~\cite{braneC}. On the other hand we 
obtain a different evolution of the  
universe with $Q \not=0$.
This universe is free from singularities of big bang and big crunch  and evolves 
with a smooth transition
between a contracting phase and an eventual expanding one. The presence of 
charge  makes it in such a way.
This provides an exotic stiff matter with  a negative energy density on the brane. 
This term dominates for small scale factor and hence induces a singular repulsive 
potential near the black hole singularity for 
the brane to be unable to access it.
 
In addition the presence of both mass $M$ and charge $Q$ induces a potential well  
which makes   bouncing and cyclic trajectories of the brane in the bulk spacetime. The key point is that the sign of 
stiff matter from $Q$ is opposite with
respect to that of radiation matter from $M$. Although this opposite sign provides  
bouncing and cyclic universes,
we immediately confront with a  difficulty that it  violates the dominant energy condition (DEC):
 $\rho \ge 0;P=\omega \rho$ with $|\omega| \le 1$~\cite{Wald}.
We remind the reader that the derivation of the DEC is based on the 4D Lorentz invariance.
 However, it was shown that  the  Lorentz invariance is apparently violated for an 
observer on the brane moving in the charged
black hole background~\cite{LorVio}.
Nowadays one may accept the negative energy density from the charge of the black hole 
as well as the violation of 
the DEC in the brane cosmology~\cite{BV2}.
 
On the other hand, the BDL approach is a genuine extension to the higher 
dimensional (Kaluga-Klein) cosmology  to take into 
account a local  matter distribution~\cite{BDL,STW,Kim}. An evolution of the brane is
 possible when the brane is fixed with respect to the
extra direction $y$ and the bulk is given by a charged AdS spacetime~\cite{CMO}. 
For this purpose we have to
introduce  negative cosmological constants and Maxwell fields for both bulk sides.
Then one finds the Friedmann equations  with two unknown  constants. 
In this case we are free to determine the corresponding mass and charge
because the mass term is usually included as an integration constant with respect to a cosmic time
on the brane and the charge is given by
an unspecified solution to the Maxwell equation under the BDL metric.
One may attempt to identify these constants
as mass and charge of the black hole if one chooses a specific bulk spacetime of two 
CTAdS black holes as did in the MDW
case. 
 
In this work, we ask mainly of whether or not bouncing and cyclic solutions for the 4D universe can be 
derived from the BDL approach.
 
The organization of this paper is as follows. In Sec.2 we review briefly the MDW 
procedure in the charged topological
AdS black holes. First we study the CFT-thermodynamics for a static brane located near 
the AdS boundary using the
AdS/CFT correspondence.  Then we will understand  the thermodynamics of a dynamic brane   
in  view of the bulk gravitational
dynamics. It is important to note  that the CFT-interpretation of radiation  and stiff matter 
is possible only when a MDW was close to the AdS boundary unless its holographic energy density is small.
Sec.3 is devoted to the BDL approach in the charged AdS background. 
For simplicity we use the same bulk stress-energy tensor
for both sides. Here we also find the same Friedmann equations with unspecified integration constants.
These can be fixed if one chooses a specific bulk spacetime like as two CTAdS black holes.
With this choice we find bouncing universes as well as cyclic universe from the BDL approach.
If not with this choice, we obtain only the bouncing universes. 
If the  exotic stiff  from charge $Q$ is considered still as an unwanted matter, we may 
answer to a question of
 how to avoid it. This may be done  by introducing an appropriate local matter on the 
brane in Sec.4. Finally we discuss our results, especially for a relation between the exotic stiff matter 
 and Lorentz
invariance violation in the brane world scenarios.

%%=======================section 2================================
\sect{MDW cosmology in the charged topological AdS black holes}
 
In general we consider an $(n+1)$-dimensional brane with  a local matter distribution
$\tau_{\mn}$ including a tension $\sigma$ which is located at between two
$(n+2)$-dimensional charged black holes.
We introduce the bulk spacetime by ${\cal M}^+$ and ${\cal M}^-$ for the positively
and negatively charged black holes and their boundaries by $\partial {\cal M}^+$ and
 $\partial {\cal M}^-$. These boundaries are fixed to coincide with the brane exactly. We keep
  the $Z_2$-symmetric geometry, although we don't need to have
 the $Z_2$-symmetry for the electric  configuration~\cite{GP}. 
Actually we have a dipole configuration ($\leftarrow$) in flat spacetime as far as the
 electric flux lines are concerned. Also we assume that there is no
 source or sink for the charge on the brane. Thus the flux lines are conserved on the brane
and thus the Israel junction condition remains unchanged.
For this purpose we start with the $(n+2)$-dimensional Einstein-Maxwell theory with
the
cosmological constant $\Lambda = - n(n+1)/2\ell^2$:
\begin{equation}
\label{2eq1}
I = \frac{1}{16\pi \g}\int_{{\cal M}^+ + {\cal M}^-} d^{n+2}x\sqrt{-g} \left ( R
-F_{MN}F^{MN} - 2 \Lambda\right),
\end{equation}
where $R$ is the curvature scalar and  $F_{MN}$ denotes the Maxwell
field\footnote{ Hereafter we use indices $M,N,\cdots$ for $(n+2)$-bulk space,
$\mu,\nu,\cdots$ for $(n+1)$-brane, and $i,j, \cdots$ for $n$-spatial space in the
brane.}. For a while we consider the right hand side $({\cal M}^+)$ of
the bulk spacetime.
Varying the action (\ref{2eq1}) yields the equations of motion
\begin{eqnarray}
\label{2eq2}
&& G_{MN} \equiv R_{MN} -\frac{1}{2}g_{MN} R
   = T_{MN}, \ \ \
T_{MN}=  2F_{MP}F_{N}^{~P}-\frac{1}{2}
 g_{MN} F^2 - \Lambda g_{MN}, \\
\label{2eq3}
&& \partial_{M}(\sqrt{-g}F^{MN})=0.
\end{eqnarray}
For completeness we need the Bianchi identity
\beq
\partial_{[M}F_{NP]}=0.
\label{2eq4}
\eeq
These equations all give us a charged topological AdS(CTAdS) black hole solution~\cite{CS,CEJM}
\begin{eqnarray}
\label{2eq5}
 && ds^2_{CTAdS} \equiv g_{MN}dx^Mdx^N= -h(r)dt^2 +h(r)^{-1}dr^2 +r^2 \gamma_{ij}dx^idx^j,  \\
\label{2eq6}
 && F_{rt}=\frac{n\omega_n}{4}\frac{Q_+}{r^n},
\end{eqnarray}
where the metric function
$h(r)$ is given by
\begin{equation}
\label{2eq7}
h(r) = k -\frac{m_+}{r^{n-1}} +\frac{q_+^2}{
   r^{2(n-1)}}
   +\frac{r^2}{\ell^2},~~m_+=\omega_n M_+,~~q_+^2=\fr{n\omega_n^2Q^2_+}{8(n-1)},
   ~~\omega_n=\frac{16 \pi G_{n+2} }{n V(M^n)}.
\end{equation}
Here $\gamma_{ij}$ is the horizon metric for a constant curvature
manifold $M^n$ with the volume $V(M^n)=\int d^nx \sqrt {\gamma}$.
The horizon geometry is thus extended to include elliptic, flat, and hyperbolic for
$k=1,0,-1$, respectively. In this sense we call it the topological black hole. 
 The event horizon $r_H$ is determined by the maximal root
of  $h(r_H)=0$. The integration constants $M_+$ and $Q_+$ can be
interpreted as the ADM  mass and electric charge of the  black hole.
 
In the spirit of the brane world scenario~\cite{RS} and  AdS/CFT correspondence~\cite{MAL}, 
let us introduce two black hole spacetimes.
We  glue it (${\cal M}^+)$ to another (${\cal M}^-$).
For simplicity we choose the same horizon geometry and
cosmological constant for ${\cal M}^+$ and ${\cal M}^-$.
 Then the thermodynamics of
 two  black holes corresponds to that for the boundary CFT with an $R$-charge
(or $Q$-potential).
We rescale the boundary metric so that it can take a form of 
$ds^2_{BCFT}=-d \tau^2+ \ell^2 \gamma_{ij}dx^idx^j$~\cite{GKPW}.
For simplicity we choose $M_+=M_-=M~(m_+=m_-=m)$, but $Q_+=-Q_-=Q/2~(q_+=-q_-=q/2)$.
As a result, we have $T_{+MN}=T_{-MN}$ for the stress-energy tensor 
\footnote{If not, in brane cosmology we have to solve
the Gauss-Godazzi equation for the non-$Z_2$ symmetric evolution by different masses and different charges.}.
This implies that a single energy $E=2E_+=2E_-$, temperature $T=T_+=T_-$, entropy $S
=2S_+=2S_-$ and pressure $p=2p_+=2p_-$ can be defined on the brane of
two boundaries using the AdS/CFT correspondence~\cite{Witten}.
Thermodynamic quantities of the corresponding CFT are given by~\cite{Cai,Noji}
\begin{eqnarray}
\label{2eq8}
&& E = 2M \fr{\ell}{r}=\frac{2\ell r_H^{n-1}}{r \omega_n}\left ( 1+\frac{r_H^2}{\ell^2}
       +\frac{n\omega_n^2 (Q/2)^2}{8(n-1)r_H^{2(n-1)}}\right), \nonumber\\
&& T =T_H \fr{\ell}{r}=\frac{\ell}{4\pi r r_H}\left((n-1) +\frac{(n+1)r_H^2}{\ell^2}
    -\frac{n\omega_n^2 (Q/2)^2}{8r_H^{2(n-1)}} \right), \nonumber \\
&& \Phi =\phi_H \fr{\ell}{r}=\frac{n \ell \omega_n Q}{8(n-1)r r_H^{n-1}}, \nonumber \\
&& S =S_{BH}= \frac{r_H^n}{2\g}V(M^n),
\end{eqnarray}
where $\Phi(=\Phi_+=-\Phi_-)$ denotes the  chemical potential conjugate
to the charge of $Q/2$.
We can rewrite the
entropy in Eq.(\ref{2eq8}) as the Cardy-Verlinde formula~\cite{Ver}
\begin{equation}
\label{2eq9}
S =\frac{2\pi r}{n}\sqrt{E_c(2(E-E_q)-E_c}),
\end{equation}
where the Casimir energy $E_c$ and electromagnetic energy $E_q$ are
given  by~\cite{CMO}
\begin{equation}
\label{2eq10}
E_c = \frac{4\ell r_H^{n-1}}{\omega_n r}, \ \
 E_q =\frac{1}{2}\Phi Q=
   \frac{n \ell \omega_n Q^2}{16(n-1)r r_H^{n-1}}.
\end{equation}
 
Now we wish to derive a dynamic equation from the moving
domain wall
approach. Actually the embedding of a moving domain wall into the black hole space  is 
a mapping of  $t \to t(\tau), r \to {\cal R}(\tau)$  with
$\dot {\cal R}^2/h({\cal R})-h({\cal R}) \dot t^2=-1$. For a good embedding, a small
black hole of $q/\ell< m/\ell<1$ is usually assumed. 
And then we obtain an induced metric $h_{\mn}$ on the brane
\beq
\label{2eq11}
ds^2_{n+1}= -d \tau^2 + {\cal R}^2(\tau) \gamma_{ij} dx^idx^j \equiv h_{\mn}dx^{\mu} dx^{\nu}.
\eeq
Here the scale factor ${\cal R}(\tau)$ will be determined
by the Israel junction condition~\cite{ISR}. For $(\partial{\cal M}_+)_{k=1,n=3}$,
the extrinsic curvature components are  given by
 
\beqa
\label{2eq12}
&&K_{\tau\tau}=K_{MN} u^M u^N =(h({\cal R}) \dot t)^{-1}(\ddot {\cal R} +h'({\cal R})
 /2)=\fr{\ddot {\cal R} +h'({\cal R})/2}
{\sqrt{\dot {\cal R}^2 +h({\cal R})}}, \\
\label{2eq13}
&&K_{\chi\chi} = K_{\theta\theta}=K_{\phi\phi}
=- h({\cal R}) \dot t {\cal R}=-\sqrt{\dot {\cal R}^2 +h({\cal R})}~{\cal R},
\eeqa
where $u^M$ is a tangent vector along the moving brane with $u^Mu_M=-1$  and
prime ($'$) stands for  derivative with respect to ${\cal R}$.
The presence of any localized matter on the brane  including a brane tension
 implies
that the extrinsic curvature jumps  across the brane. At this stage let us
glue it to another bulk of ${\cal M}_-$ with the same ${\cal
R}(\tau)$ but an opposite embedding of $K_{-\mn}=-K_{+\mn}(=-K_{\mn})$.
This jump is then  described  by the Israel junction
condition\footnote{The Gibbons-Hawking term $\fr{1}{\kappa^2} \int d^{n+1} x \sqrt{h} K $
 and the Hawking-Ross term $\fr{2}{\kappa^2} \int d^{n+1} x \sqrt{h} F^{MN}n_M A_N$
  are necessary for
embedding of the brane into the
charged black hole. However, the electric flux across the brane is conserved 
and thus the Israel junction condition remains unchanged even for $F_{MN} \not=0$~\cite{GP}. Here we no longer consider  one-sided brane
cosmology. Hence we keep the $Z_2$-symmetry which was
important  in two-sided brane cosmology.}
\beq
\label{2eq14}
K_{+\mn}-K_{-\mn}=-\kappa^2 \left(
\tau_{\mu\nu}-\fr{1}{n}\tau^{\lambda}_{~\lambda}h_{\mu\nu} \right)
\eeq
with  the bulk gravitational constant $\kappa^2=8 \pi G_{n+2}$.
The 4D perfect fluid is introduced for a local stress-energy tensor on the
brane
\beq
\tau_{\mu \nu}=(\varrho +p)u_{\mu}u_{\nu}+p\:h_{\mu\nu}.
\label{2eq15}
\eeq
Here $\varrho=\rho_l+ \sigma$ $(p=P_l-\sigma)$, where $\rho_l $ $(P_l)$
are the energy density (pressure)
of a  local matter and $\sigma$ is a brane tension.
In the case of $\rho_l=P_l=0$, the right hand side  of
Eq.~(\ref{2eq14}) leads to  a form of the Randall-Sundrum  case as $-\fr{\sigma \kappa^2}{n}
h_{\mu\nu}$~\cite{RS}.
The spatial  components of the junction condition (\ref{2eq14}) lead to
 
\beq
\sqrt{h({\cal R}) + \dot {\cal R}^2}=\fr{\kappa^2}{2n}\sigma {\cal R}.
\label{2eq16}
\eeq
In two black hole spacetimes,
we use the  tension $\sigma=2n/(\kappa^2\ell)$ to obtain a critical brane.
The above equation    leads to the first Friedmann equation
\beq
H^2=- \fr{k}{{\cal R}^2} +\fr{m}{{\cal R}^{n+1}} -\fr{q^2}{{\cal R}^{2n}}.
\label{2eq17}
\eeq
Its time rate as the second Friedmann equation is given by
 
\beq
\dot H= \fr{k}{{\cal R}^2} -\fr{n+1}{2}\fr{m}{{\cal R}^{n+1}} +  \fr{nq^2}{{\cal R}^{2n}}.
\label{2eq18}
\eeq
Introducing a
CFT-energy density $\rho_{r}= E/V$ with $ E=2M \ell/{\cal R}$, its pressure $P_{r}=\rho_{r}/n$, a charge
density $\rho_Q=Q/V$ , and its electric potential $\Phi=n \ell \omega_n Q/8(n-1){\cal R}^n$
with the volume of the brane $V={\cal R}^n
V(M^n)$,  two Friedmann equations take the forms
\beqa
\label{2eq19}
&& H^2=- \fr{k}{{\cal R}^2} +\fr{16 \pi G_{n+1}} {n(n-1)} \Big(\rho_{r}
 -\fr{1}{2} \Phi \rho_Q \Big), \\
\label{2eq20}
&& \dot H= \fr{k}{{\cal R}^2} -\fr{8 \pi G_{n+1}} {n-1} \Big(\rho_{r} +P_{r} -
\Phi \rho_Q \Big).
\eeqa
In deriving these, we use an important relation for the two-sided brane scenario
\beq
\g=\fr {2\ell}{n-1} G_{n+1}
\label{2eq21}
\eeq
and the conservation law of $\dot \rho =-n H (\rho+P)$.
This means that the cosmological evolution is derived by the
energy density and  pressure of the CFT-radiation matter plus
those of the electric potential energy. Eqs. (\ref{2eq19}) and (\ref{2eq20}) are identified with those of the one-sided brane world
~\cite{Youm} because two black hole masses and charges on both sides are combined with
Eq.(\ref{2eq21}) to give the same ones.   These equations
can be further  rewritten as the cosmological Cardy-Verlinde formula and  a defining relation for
the Bekenstein-Hawking energy as
\beqa
\label{2eq22}
&& S_H= \fr{2 \pi {\cal R}}{n} \sqrt{ E_{BH}[2(E- \Phi Q/2)-k
E_{BH}]},\\
\label{2eq23}
&& kE_{BH}=n(E+pV-\Phi Q-{\cal T}_H S_H).
\eeqa
Here the Hubble temperature ${\cal T}_H=-\fr{\dot H}{ 2 \pi H}$ is expressed in
terms of the Hubble parameter and its time rate only.
 
Another interpretation is also possible for the Friedmann equations on the brane if one
defines  a  charge stiff density $\rho_{csti}=\Phi \rho_Q/2 $ with the equation of state :
$P_{csti}=\rho_{csti}$~\cite{Myung1}. Then one finds
\beqa
\label{2eq24}
&& H^2=- \fr{k}{{\cal R}^2} +\fr{16 \pi G_{n+1}} {n(n-1)} \rho_h, \\
\label{2eq25}
&& \dot H= \fr{k}{{\cal R}^2} -\fr{8 \pi G_{n+1}} {n-1}
\Big(\rho_h +P_h\Big),
\eeqa
where a holographic energy density $\rho_h$ and its pressure $P_h$ are defined by
\beq
\rho_h=\rho_{r}-\rho_{csti},~~~P_h=P_{r}-P_{csti}=\fr{\rho_r}{n}-\rho_{csti}.
\label{2eq26}
\eeq
This means that the mass of the black hole gives a CFT-radiation matter
holographically, whereas the charge of the black hole induces an exotic
CFT-stiff matter on the moving domain wall because the sign in the front of $\rho_{csti}$
is negative.
Thus it is shown that the bulk gravitational dynamics   leads to the standard cosmology with
a radiation matter of $\rho_{r} \sim {\cal R}^{-(n+1)}$ and
a stiff matter of $\rho_{csti} \sim {\cal R}^{-2n}$ on the brane.
However,
this CFT-interpretation on the brane
is always not true. This is possible only  when the brane is  located at
large distance from the center of AdS space
unless the holographic energy density $\rho_h$  is  small~\cite{GP}.
Applying the Hamiltonian approach to calculation of  the exact
energy on the brane, the energy density $\rho_e$ and its pressure $P_e$ measured by an
observer on the brane is given by
\beqa
\label{2eq27}
&&\rho_e\equiv  \fr{2n}{\kappa^2 \ell} \xi({\cal R})=\fr{2n}{\kappa^2 \ell}
 \Big[ \sqrt{1+\fr{\ell^2}{
{\cal R}^2} \Big(\fr{m}{{\cal R}^{n-1}} -\fr{q^2}{{\cal R}^{2n-2}}
\Big)}-1\Big],\\
\label{2eq28}
&&P_e=-\rho_f + \fr{\ell}{\kappa^2 \xi({\cal R})}
\Big[\fr{(n+1)m}{{\cal R}^{n+1}} -\fr{2n q^2}{{\cal R}^{2n}}
\Big].
\eeqa
Here it is not guaranteed that $\rho_f$ is always positive.
The Friedmann equations lead to those of the conventional brane cosmology
 
\beqa
\label{2eq29}
&& H^2=- \fr{k}{{\cal R}^2} +\fr{16 \pi G_{n+1}} {n(n-1)} \rho_e +
\fr{\kappa^4}{4n^2} \rho^2_e, \\
\label{2eq30}
&& \dot H= \fr{k}{{\cal R}^2} -\fr{8 \pi G_{n+1}} {n-1}
\Big(\rho_e +P_e\Big) -  \fr{\kappa^4}{4n} \rho_e(\rho_e+P_e).
\eeqa
Small $\rho_e$  means either that the brane is close to the AdS boundary:
\beq
\label{2eq31}
\lim_{{\cal R} \to
  \infty} \fr{\ell^2}{
{\cal R}^2} \Big(\fr{m}{{\cal R}^{n-1}} -\fr{q^2}{{\cal R}^{2n-2}} \Big)\to
{\rm small}
\eeq
or that a holographic term  is small
 
\beq
\label{2eq32}
\Big(\fr{m}{{\cal R}^{n-1}} -\fr{q^2}{{\cal R}^{2n-2}} \Big) \to
{\rm small}
\eeq
which implies a small black hole of $q/\ell<m/\ell<1$.
In this approximation the  energy density is given by
\beq
\rho_e \approx
\fr{n \ell}{\kappa^2} \Big(\fr{m}{{\cal R}^{n+1}} -\fr{q^2}{{\cal R}^{2n}}
\Big)=\rho_r-\rho_{csti} =\rho_h
\label{2eq33}
\eeq
and for its pressure
\beq
P_e \approx \fr{\rho_r}{n} -\rho_{csti} =P_r -P_{csti}=P_h.
\label{2eq34}
\eeq
Neglecting $\rho_e^2$ and $\rho_e P_e$ terms in (\ref{2eq29}) and (\ref{2eq30}),
 we recover our equations (\ref{2eq24}) and (\ref{2eq25}).
This means that our  holographic interpretation of
the cosmological Cardy-Verlinde formula for  Eq.(\ref{2eq22}) has
limitation and approximation because it is based on an approximate
equation (\ref{2eq8}). Unless  we  assume small $\rho_e$,
 a lot of terms are generated in the Friedmann equations.
In this case  we may establish the unconventional duality, and the
holographic duality corresponds to   an
approximation of the unconventional duality.
 
%%====================section 3=============================
 
\sect{BDL cosmology in the charged background}
 
In the MDW approach the bulk space time is always fixed.
This means that we can determine the gravitational dynamics of the brane
exactly from the given bulk black hole background. Actually, we
don't know precisely what kind of  higher dimensional  theory
we should start with, even if string theories provide us some information about it.
 In this sense the moving domain wall method is considered as a very restrictive approach  
  although it provides us an exact evolution of the brane.
 The BDL approach is a genuine extension to the
Kaluza-Klein
cosmology to account for a  local matter distribution on the brane.
Hence it is very desirable to study the BDL brane cosmology in the
charged background.
Following~\cite{BDL}, we assume the Gaussian-normal metric
for an ($n+2$)-dimensional spacetime
\begin{equation}
\label{3eq1}
ds^2_{BDL} = -c^2(t,y) dt^2 +a^2(t,y)\gamma_{ij}dx^idx^j +b^2(t,y)dy^2,
\end{equation}
where $\gamma_{ij}$ is the previous metric of an $n$-dimensional space with constant
curvature $n(n-1)k$. In the orthogonal basis, we express the
Einstein tensor $G_{\hat M \hat N}$ in Eq.(\ref{2eq2}) in terms of
the BDL metric as
 
\begin{eqnarray}
\label{3eq2}
G_{\hat t \hat t} &=& n \left [\frac{\dot a}{ac^2}\left(\frac{n-1}{2}
    \frac{\dot a}{a}+\frac{\dot b}{b}\right)
   -\frac{1}{b^2}\left (\frac{a''}{a}+\frac{a'}{a}\left( \frac{n-1}{2}
    \frac{a'}{a}-\frac{b'}{b}\right) \right)
   +\frac{n-1}{2}\frac{k}{a^2}\right ], \nonumber \\
G_{\hat y \hat y} &=&n \left [ \frac{a'}{ab^2}\left(
   \frac{n-1}{2}\frac{a'}{a} +\frac{c'}{c} \right)
    -\frac{1}{c^2}\left( \frac{\ddot a}{a}
    +\frac{\dot a}{a}\left ( \frac{n-1}{2}\frac{\dot a}{a}-\frac{\dot c}{c}
    \right)\right) -\frac{n-1}{2}\frac{k}{a^2}\right], \nonumber \\
G_{\hat t \hat y} &=& n \left (\frac{\dot a c'}{abc^2}
                  + \frac{a'\dot b}{ab^2 c} -\frac{\dot a'}{abc} \right),
      \nonumber \\
G_{\hat i \hat j} &=& \frac{\delta_{ij}}{b^2} \left[ (n-1)\frac{a''}{a}
          +\frac{c''}{c}+\frac{n-1}{2}\frac{a'}{a}
            \left((n-2)\frac{a'}{a} +2\frac{c'}{c}
           \right) -\frac{b'}{b}\left ((n-1)\frac{a'}{a}
        +\frac{c'}{c}\right) \right] \nonumber \\
     & & + \frac{\delta_{ij}}{c^2}\left [-(n-1)\frac{\ddot a}{a}
     -\frac{\ddot b}{b} +\frac{\dot b}{b}\left (\frac{\dot c}{c}- (n-1)
      \frac{\dot a}{a}\right)
     +\frac{n-1}{2} \frac{\dot a}{a} \left (2 \frac{\dot c}{c} -(n-2)
         \frac{\dot a}{a}\right) \right] \nonumber \\
    & & -\frac{(n-1)(n-2)}{2}\frac{k}{a^2}\delta_{ij},
\end{eqnarray}
where the dot (prime) stand for differentiation with respect to $t$ ($y$).
Suppose that an $(n+1)$-dimensional brane is located at $y=0$. In
the two-sided brane world, the stress-energy tensors are given by~(\ref{2eq2}),
but we don't need to choose the same initially.
Furthermore $T_{\hat t \hat y}=0$ implies that
there is no local matter flow along the extra direction.
A local stress-energy
tensor  including a  brane  tension $\sigma$ is assumed to be 
\begin{equation}
\label{3eq3}
\tilde \tau_{\mu}^{\ \nu} = \frac{\delta(y)}{b}{\rm
diag}(-\varrho,p,\cdots,p,0).
\end{equation}
Let us denote the gap of a given function $f$ at $y=0$ by
$[f]=f(0_+) -f(0_-)$ and its average by $\{f\}=(f(0_+)
+f(0_-))/2$,
where $+(-)$ denote $y>0(y<0)$, respectively.
The functions $a$, $b$, $c$ in the metric~(\ref{3eq1}) are continuous
at $y=0$, but their derivatives are not. So the second derivative
takes the form~\cite{STW}
\begin{equation}
\label{3eq4}
f''=f''|_{(y\ne 0)} +[f']\delta(y).
\end{equation}
It is then straightforward to write down the gaps in  ($\hat t \hat t$), ($\hat y \hat y$)
and ($\hat i \hat j$) components of the Einstein equation~(\ref{2eq2}), respectively:
\begin{eqnarray}
\label{3eq5}
&& \frac{n}{b^2_0}\left (- (n-1)\frac{[a']\{a'\}}{a^2_0}
   +\frac{[a']\{b'\}}{a_0b_0} +\frac{\{a'\}[b']}{a_0b_0}\right)
 = T_{\hat t \hat t}(0_+) -T_{\hat t \hat t}(0_-),
     \nonumber \\
&& \frac{n}{b_0^2}\left ( (n-1)\frac{[a']\{a'\}}{a_0^2}
    +\frac{[a']\{c'\}}{a_0c_0} +\frac{\{a'\}[c']}{a_0c_0}
     \right) = T_{\hat y \hat y}(0_+) -T_{\hat y \hat y}(0_-),
   \nonumber \\
&& \frac{(n-1)}{b_0^2}\delta_{ij} \left( (n-2)\frac{[a']\{a'\}}{a_0^2}
   +\frac{[a']\{c'\}}{a_0c_0} +\frac{\{a'\}[c']}{a_0c_0}
   -\frac{[a']\{b'\}}{a_0b_0}-\frac{\{a'\}[b']}{a_0b_0} \right. \nonumber \\
&& ~~~~~~~~~~~~\left. -\frac{1}{n-1}\frac{[b']\{c'\}}{b_0c_0}
     -\frac{1}{n-1}\frac{\{b'\}[c']}{b_0c_0} \right )=T_{\hat i \hat j}(0_+)
   -T_{\hat i \hat j}(0_-).
\end{eqnarray}
where  quantities with subscript ``0" denote those at $y=0$. The
$\delta(y)$-function parts in  ($\hat t \hat t$) and ($\hat i \hat j$)
components of the Einstein equation with Eq.(\ref{3eq3}) lead to the Israel junction condition
 
\begin{equation}
\label{3eq6}
\frac{[a']}{a_0b_0}=-\frac{\kappa^2}{n}\varrho, \ \  \  \frac{[c']}{b_0c_0}
   =\kappa^2\left (p +\frac{n-1}{n}\varrho\right).
\end{equation}
 
On the other hand, the average part of  ($\hat y \hat y$) component is given by
 
\begin{eqnarray}
\label{3eq7}
&& \frac{1}{c_0^2}\left (\frac{\ddot a_0}{a_0} +\frac{\dot a_0}{a_0}
   \left(\frac{n-1}{2} \frac{\dot a_0}{a_0} -\frac{\dot c_0}{c_0}\right)
   \right)
=-\frac{1}{2n}\left (T_{\hat y\hat y}(0_+) +
        T_{\hat y \hat y}(0_-)\right)   \nonumber \\
 && ~~~~~ +\frac{n-1}{2}\left (-\frac{k}{a^2_0}
    +\frac{1}{4}\left (\frac{[a']}{a_0b_0}\right)^2
   +\left(\frac{\{a'\}}{a_0b_0}\right)^2 \right)
   +\frac{1}{4} \frac{[a'][c']}{a_0b_0^2 c_0}
  +\frac{\{a'\}\{c'\}}{a_0b_0^2 c_0}.
\end{eqnarray}
The Maxwell equation~(\ref{2eq3}) and the Bianchi identity
 (\ref{2eq4}) under the BDL metric~(\ref{3eq1})
have a solution
 
\begin{equation}
\label{3eq8}
F_{yt} =\frac{{\cal Q}bc}{a^n},
\end{equation}
where ${\cal Q}$ is an unknown integration constant. Thus the  stress-energy
tensor including the Maxwell field $F_{MN}$ and cosmological constant $\Lambda$ is given by
\begin{equation}
\label{3eq9}
T_{\hat M}^{~\hat N} ={\rm diag}\left (\frac{n(n+1)}{2\ell^2}
  -\frac{{\cal Q}^2}{a^{2n}}, \frac{n(n+1)}{2\ell^2} +\frac{{\cal Q}^2}{a^{2n}},
   \cdots, \frac{n(n+1)}{2\ell^2} +\frac{{\cal Q}^2}{a^{2n}},
   \frac{n(n+1)}{2\ell^2}-\frac{{\cal Q}^2}{a^{2n}} \right).
\end{equation}
In order to go  parallel with the moving domain wall approach,
 we consider a simple case in which  the
bulk stress-energy tensors are identical on two sides of the brane and thus the bulk is $Z_2$-symmetric.
This is the same
situation as was considered  in Sec. 2. One
then has $\{f'\}=0$ for $f=a,b,c$. Also $G_{\hat t \hat y}=0$ is satisfied if
$c(t,y)=\dot a(t,y)/\dot a_0$ with $b(t,0)=1$. This means that the position of the brane
is always fixed along the extra direction during whole evolution. 
This allows us to set $c_0 \equiv c(t,0)=1$ so that $d\tau \equiv c(t,0)dt=dt$. 
The Hubble parameter $H$ on the brane  is defined as
$H =\dot a_0/a_0 \equiv \dot {\cal R}/{\cal R}$.
The left hand side of Eq.~(\ref{3eq7}) can be rewritten as
\begin{equation}
\label{3eq10}
\frac{1}{c_0^2}\left (\frac{\ddot a_0}{a_0} +\frac{\dot a_0}{a_0}
   \left(\frac{n-1}{2} \frac{\dot a_0}{a_0} -\frac{\dot c_0}{c_0}\right)
   \right) = \frac{1}{2{\cal R}^n}\frac{d}{d{\cal R}} \Big(H^2{\cal R}^{n+1} \Big),
\end{equation}
while the right hand side of Eq.~(\ref{3eq7})  can be calculated with  Eqs.~(\ref{3eq6}) and
(\ref{3eq9}) to yield
\begin{equation}
\label{3eq11}
\frac{1}{2{\cal R}^n}\frac{d}{d{\cal R}} \Big( H^2{\cal R }^{n+1} \Big)=
-\frac{\kappa^4 }{8n^2}\varrho\left(
     2n p +(n-1)\varrho \right) -\frac{n+1}{2\ell^2}
    -\frac{n-1}{2}\frac{k}{{\cal R}^2} +\frac{{\cal Q}^2}{n {\cal R}^{2n}}.
\end{equation}
In general we include  a  local matter distribution with $\tau_{\mn}$ on the brane.
In this case we have $\varrho =\rho_l + \sigma,~p=P_l-\sigma$ 
with $P_l=\varpi \rho_l,~\sigma =2n/\kappa^2 \ell$.
Then the brane becomes a critical one and 
Eq.~(\ref{3eq11}) reduces to
\beq
\label{3eq12}
\frac{d}{d{\cal R}} \Big(H^2 {\cal R}^{n+1} \Big)=
-(n-1)k {\cal R}^{n-2} +
\frac{2{\cal Q}^2}{n{\cal R}^n} 
-\fr{\kappa^2(n \varpi-1){\cal R}^n}{n\ell}\rho_l
-\fr{\kappa^4}{4n^2}(2n \varpi + n-1){\cal R}^n \rho_l^2.
\eeq
Assuming $\rho_l \sim {\cal R}^{-n(1+\varpi)}$, one integrates this equation
to yield
\beq
\label{3eq13}
H^2 =-\frac{k}{{\cal R}^2} +\frac{{\cal C}}{{\cal R}^{n+1}} -\frac{2}{n(n-1)}
\frac{{\cal Q}^2}{{\cal R}^{2n}}+ \fr{16 \pi G_{n+1}}{n(n-1)} \rho_l + \fr{\kappa^4}{4n^2}
\rho_l^2,
\eeq
where ${\cal C}$ and ${\cal Q}$ are  two integration constants~\cite{BDL,CMO}.
 This governs an evolution of the fixed  brane  sandwiched in the two charged AdS spacetimes.

For $\rho_l=P_l=0$ case, let us  compare our equation (\ref{3eq13}) with  (\ref{2eq17}) of
the moving domain wall in the CTAdS black holes which can be rewritten further as
\begin{equation}
\label{3eq14}
H^2 =-\frac{k}{{\cal R}^2} +\frac{\omega_n M}{{\cal R}^{n+1}}
 -\frac{n\omega^2_n Q^2}{32(n-1){\cal R}^{2n}}.
\end{equation}
By analogy, we rewrite two constants as
\begin{equation}
\label{3eq15}
{\cal C}= \omega_n  \tilde M, \ \ \ \ {\cal Q} = \pm \frac{n \omega_n }{4} \fr{\tilde Q}{2}.
\end{equation}
Our equation (\ref{3eq13}) exactly coincides with Eq.~(\ref{3eq14}) if $M=\tilde M$ and $Q=\tilde Q$. 
In case of  the moving brane Eq.~(\ref{3eq14}), the bulk is just fixed as    two CTAdS  black holes with
the same mass $M$ and   charge $Q/2$. These two parameters encode 
all information of the bulk and are used to
describe a strongly coupled CFT  on the brane
using the finite-temperature AdS/CFT correspondence.
On the other hand, in deriving Eq.~(\ref{3eq13}), we don't  know
 what  the precise AdS geometry is.  But we   include
 the bulk Maxwell fields.  Certainly, the two integration constants
${\cal C}$ and ${\cal Q}$ will carry with information of the
bulk, as the initial conditions.
 This  shows that a holographic duality may be realized within the BDL approach~\cite{Kim}.

Now we are position to obtain a solution to equation (\ref{3eq13}).
As an explicit computation we choose a case of  $n=3, \rho_l=P_l=0$. 
Introducing  a conformal time $\eta$, 
defined by $d\tau={\cal R}(\eta)d \eta$, one finds a solution for $k=1$ closed geometry
\beq
\label{3eq16}
{\cal R}^{k=1}(\eta)= \sqrt{\fr{\omega_3 \tilde M}{2} \Big( 1- \tilde c_1 \cos (2 \eta) \Big)}, 
~~\tilde c_1=\sqrt {1- \fr{3 \tilde Q^2}{16 \tilde M^2}}.
\eeq
Here we do not guarantee the reality of  $\tilde c_1$: $\tilde Q /2< 2 \tilde M/\sqrt{3}$,
approximately, a small black hole condition of $q/\ell<m/\ell<1$.
because  the values of two parameters $ \tilde M$ and $\tilde Q$ are  arbitrary.
This means that $\tilde c_1$ may become a pure imaginary.  On the other hand, for the MDW case 
the reality condition of $\tilde c_1$ comes from the condition for the existence 
of the event horizon  in the 5D AdS Reissner-Nordstrom (AdSRN$_5$) black hole~\cite{MP}. 
Hence a cyclic solution is   allowed.
Then the universe evolves periodically between a maximal distance ${\cal R}_{max}$ and 
a minimal distance ${\cal R}_{min}$ with 
\beq
\label{3eq17}
{\cal R}^{k=1}_{max/min}= \sqrt{\fr{\omega_3  M}{2} ( 1\pm c_1)}, 
~~c_1=\sqrt {1- \fr{3  Q^2}{16 M^2}}~~>0.
\eeq
If one chooses $\tilde M=M$ and $\tilde Q=Q$, then we find the above bouncing universe.
However, this case rarely occurs in the BDL approach because  the bulk spacetime should be
chosen specially as two AdSRN$_5$ black holes.
 
For $k=-1$ open universe, one has the solution
\beq
\label{3eq18}
{\cal R}^{k=-1}(\eta)= \sqrt{\fr{\omega_3 \tilde M}{2} \Big(\tilde c_2 \cosh (2 \eta)-1 \Big)}, 
~~\tilde c_2=\sqrt {1+ \fr{3 \tilde Q^2}{16 \tilde M^2}}~~>1.
\eeq
This is acceptable because $c_2$ is always real. The brane is initially contracting and then bounces to
an expansion. At the turning point of $\eta=0$, the minimal radius takes the form
\beq
\label{3eq19}
{\cal R}^{k=-1}_{min}= \sqrt{\fr{\omega_3 \tilde M}{2} (c_2-1)}. 
\eeq
In the absence of the Maxwell fields ($\tilde Q=0$), one immediately finds ${\cal R}^{k=-1}_{min}=0$.
 
In the case of $k=0$ flat universe we have 
\beq
\label{3eq20}
{\cal R}^{k=0}(\eta)= \sqrt{ \fr{3 \tilde Q^2 \omega_3}{64 \tilde M} + \omega_3 \tilde M \eta^2}.
\eeq
 
Also we have a bouncing universe with the minimal distance 
${\cal R}^{k=0}_{min}= \sqrt{3 \tilde Q^2 \omega_3/64 \tilde M}$ at the turning point $\eta=0$.
Obviously this goes to zero in the limit of $\tilde Q \to 0$.
Hence we find bouncing universes for $k=-1,0$ cases from the BDL approach.
The cyclic closed universe is only allowable if the bulk spacetime are chosen specially as   
two AdSRN$_5$ black holes.
In this case, all solutions remain fixed with respect to the extra direction $y$.
However, there is no restriction on $\tilde M$ and $\tilde Q$, for example, a small black hole condition of
$\tilde Q /2< 2 \tilde M/\sqrt{3}$ required in the MDW solutions.

\sect{Brane cosmologies with a local matter}
\subsection{MDW approach}
 
First of all  we wish to remark the sign of $\rho_h=\rho_r-\rho_{csti}$ in
Eq.(\ref{2eq26}).
At the very early universe we have a negative holographic energy density
of $\rho_h<0$ because $\rho_{csti}$ dominates.  On the other hand,
in the standard cosmology one usually uses the
dominant energy condition (DEC) which postulates that the local energy
density is non-negative for all observers :$\rho \ge 0, P=\omega \rho$
with $|\omega|\le 1$. But we do not maintain this condition because of $\rho_h < 0$ at 
the very early universe.
In this section  we may deal with this problem by introducing an appropriate local matter
distribution on the brane~\cite{Myung1}.
In the case of $\rho_l \neq 0$ with $P_l=\varpi \rho_l$, from Eqs.(\ref{2eq14}) and (\ref{2eq15})
we obtain two Friedmann equations 
\beqa
\label{4eq1}
&& H^2=- \fr{k}{{\cal R}^2} +\fr{16 \pi G_{n+1}} {n(n-1)}
\Big(\rho_r -\rho_{csti} + \rho_l \Big) + \fr{\kappa^4}{4n^2} \rho_l^2, \\
\label{4eq2}
&& \dot H= \fr{k}{{\cal R}^2} -\fr{8 \pi G_{n+1}} {n-1}
\Big(\rho_r -\rho_{csti}+\rho_l +P_r -P_{csti}  +P_l\Big) -n\fr{\kappa^4}{4n^2}
\rho_l(\rho_l+P_l).
\eeqa
If a local stiff matter with $P_{lsti}=\rho_{lsti}>\rho_{csti}$ is chosen, we  achieve the
positivity of the  total energy density $\rho_T= \rho_h+ \rho_{lsti} =\rho_r +
(\rho_{lsti}-\rho_{csti})>0 $ in  Eq.(\ref{4eq1}).
Even though we avoid an exotic stiff-matter problem thanks to a local
stiff matter, it gives rise to a higher-order term
of $\rho_{lsti}^2 \sim {\cal R}^{-4n}$, which have not been found in the
standard and brane cosmologies. If this exists, it will contribute to the very early universe.
 
In order to find another case, we rewrite the above equations using $\rho^2_{cdust}=Q^2/V^2$ as
\beqa
\label{4eq3}
&& H^2=- \fr{k}{{\cal R}^2} +\fr{16 \pi G_{n+1}} {n(n-1)}
\Big(\rho_r + \rho_l \Big) + \fr{\kappa^4}{4n^2} \Big(\rho_l^2-\rho_{cdust}^2 \Big), \\
\label{4eq4}
&& \dot H= \fr{k}{{\cal R}^2} -\fr{8 \pi G_{n+1}} {n-1}
\Big(\rho_r +P_r  +P_l\Big) -n\fr{\kappa^4}{4n^2}
\Big[\rho_l(\rho_l+P_l) -\rho^2_{cdust} \Big].
\eeqa
Including a local dust matter with
$P_{ldust}=0,~\rho^2_{ldust}>\rho^2_{cdust}$, then we maintain the
positivity of the last term in Eq.(\ref{4eq3}). However, we obtain
an effective energy density which is composed of two different matters
: a CFT-radiation matter with $\rho_r\sim {\cal R}^{-(n+1)}$ plus
a local dust matter with $\rho_{ldust} \sim {\cal R}^{-n}$.
 
\subsection{BDL approach}
 
Considering Eq.~(\ref{3eq13}) together with (\ref{3eq15}), 
we can lead to Eqs.(\ref{4eq1}) and (\ref{4eq2}) but replacing $\rho_r, \rho_{csti}$ by
 $\tilde \rho_r, \tilde \rho_{csti}$ of $\tilde M, \tilde Q^2$.
Also another case  of Eqs.(\ref{4eq3}) and (\ref{4eq4})
can be recovered from the BDL approach by replacing $\tilde \rho_r, \tilde \rho^2_{cdust}$ of
 $\tilde M, \tilde Q^2$. Hence we may avoid an exotic
stiff matter due to the bulk  Maxwell field by inserting an
appropriate local matter on the brane within the BDL scheme.
In addition, we have  degrees of freedom to make the positivity of the total energy density
because there are no restrictions on determining $\tilde M$ and $ \tilde Q$ in the BDL approach.

\sect{Discussions}
A moving domain wall  in the charged topological AdS black holes provides bouncing and cyclic
universes. This solution is possible because  an exotic stiff matter term arises from the bulk  
charge $Q$ of the black hole 
in a holographic way. On the other hand this term obviously violates the dominant energy 
condition of keeping the positive
energy density for all observers on the brane. In the BDL approach we confirm from 
Eq.(\ref{3eq13}) that the  Maxwell field with a charge $ \tilde {\cal Q}$ generates  an exotic 
stiff matter with a negative sign. 
As was reported in~\cite{LorVio},
one cannot guarantee the positivity of holographic energy densities. But there is no problem in 
any local matter on the brane because the 4D Lorentz invariance is always maintained.  
 
The embedding of a moving domain wall into the black hole background is just the 2$\to$1 mapping
of $t \to t(\tau),~r \to {\cal R}(\tau)$ with a timelike tangent vector $u^M=(\dot t, \dot {\cal R},0,0,0),
~u^Mu_M=-1$ 
and a spacelike normal vector $n_M=(-\dot {\cal R}, \dot t,0,0,0),~n^Mn_M=1,~u^Mn_M=0$. 
These  lead to
only one condition of $-h({\cal R}) \dot t^2+\dot {\cal R}^2/h({\cal R})=-1$ which makes the 
timelike brane during whole
evolution~\cite{Myung2}. However, it is well known that for a timelike bulk 
coordinate $t=t(\tau)$ and spacelike bulk coordinate $r={\cal R}(\tau)$, their role is 
exchanged into each other when the brane crosses the horizon of the black hole 
at ${\cal R}=r_H$ with $h({\cal R})=0$. This means that the induced 
metric (\ref{2eq11}) will have  differently defined Lorentz symmetry at different 
points along the radial timelike geodesics
 to maintain a speed of light $c=1$. 
This implies that the bulk spacetime of (\ref{2eq5}) globally violates 4D Lorentz invariance, 
leading to apparent
violations of Lorentz invariance in view of an observer on the brane due to the bulk 
gravitational effects.
An important fact of this consequence is that the speed of gravitational waves which 
propagates through
the bulk spacetime would be different from the speed of light waves which propagates 
along the brane.
Explicitly the gravitational wave will arrive at the brane faster than the light 
wave~\cite{LorVio}. 
This could cause apparent violations of the causality from the view of an observer on the 4D brane.
On the other hand the  dominant energy condition  usually requires the causality on the 4D spacetime.
Hence it arrives that we don't need to require the dominant energy condition on the 
holographic energy density
 of $\rho_h=\rho_r -\rho_{csti}$ from the bulk configuration. The mass of the 
black hole happens to give a positive
energy density, whereas the charge happens to be a negative energy density. 
This is considered as a result of the nature of the bulk configuration. 
 
In the BDL approach we note first that two metrics (\ref{2eq5}) and (\ref{3eq1}) are 
equivalent even for the charged black hole background~\cite{MSM,LorVio}. 
Hence we expect that the dominant energy condition is not necessarily required 
on the holographic energy density
 of $\tilde \rho_h=\tilde \rho_r -\tilde \rho_{csti}$ from the  Maxwell fields.
Our result of  Eq.(\ref{3eq13})  shows an explicit example for this direction. 
 
In conclusion we obtain the bouncing universes from the BDL approach in the 
charged bulk spacetime.
This method provides a more general approach to study the brane cosmology 
than the moving domain wall.
Especially considering a special bulk of two AdSNR black holes, then we can 
find a cyclic universe for  $k=1$ closed geometry.
These solutions arise mainly from an exotic stiff matter term originated from the  
Maxwell field.
If one wishes to avoid this exotic matter on the brane, we may include an appropriate 
local matter. 
In this case we may use this model to describe a core of neutron stars consistent 
with causality~\cite{BFKMP}

\section*{Acknowledgments}
We thank Henry Tye for helpful discussions.
This work was supported in part by  KOSEF, Project Numbers. 2000-1-11200-001-3
and R02-2002-000-00028-0.
 
%\newpage

%%=======================references===========================


\begin{thebibliography}{99}
 
\bibitem{KOST} J. Koury, B. A. Ovrut, P. J. Steinhardt and N. Turok, 
Phys. Rev. {\bf D64}, 123522 (2001) [hep-th/0103239].
 
\bibitem{KOSST} J. Koury, B. A. Ovrut, N. Seiberg, P. J. Steinhardt and 
N. Turok, Phys. Rev. {\bf D65}, 086007 (2002)
                [hep-th/0108187].
 
\bibitem{MP} S. Mukherji and M. Peloso, {\it Bouncing and cyclic universes 
from brane models}, hep-th/0205180.
 
\bibitem{BS} Ph. Brax and D. A. Steer, {\it A comment on the bouncing and cyclic 
branes in more than one 
        extra-dimension}, hep-th/0207280.
\bibitem{CR} H. Chamblin and H. Reall, Nucl. Phys. {\bf B562}, 133 (1999) [hep-th/9903225].
 
\bibitem{Kraus}P. Kraus, JHEP {\bf 9912}, 011 (1999) [hep-th/9910149].
 
\bibitem{Ida}D. Ida, JHEP {\bf 0009}, 014 (2000) [gr-qc/9912002].
 
\bibitem{BCG} P. Bowcock, C. Charmousis, and R. Gregory, Class. 
Quant. Grav. {\bf 17}, 4745 (2000) [hep-th/0007177].
\bibitem{MSM}S. Mukohyama, T. Shiromizu and K. Maeda, Phys.Rev. {\bf D63}, 029901 (2001)
                          [ hep-th/9912287].
 
\bibitem{CEJM} A. Chamblin, R. Emparan, C. V. Johnson, and R. C. Myers, 
Phys. Rev. {\bf D60}, 064018 (1999)
               [hep-th/9902170].
 
\bibitem{CS} R. G. Cai and K. S. Soh, Phys. Rev. {\bf D59}, 044013 (1998) [gr-qc/9808067].
 
\bibitem{BV1} C. Barcelo and M. Visser, Phys. Lett. {\bf B482}, 183 (2000) [hep-th/0004056].
 
\bibitem{BM} A.K. Biswas and S. Mukherji, JHEP {\bf 0103}, 046 (2001) [hep-th/0102138].
 
\bibitem{Myung1} Y.S. Myung, {\it Standard cosmology from the brane 
cosmology with a localized matter}, hep-th/0103241.
 
\bibitem{CMO} R.G. Cai, Y. S. Myung and N. Ohta, Class. Quant. 
Grav. {\bf 18}, 5429 (2001) [hep-th/0105070].
 
\bibitem{Youm} D. Youm, Mod. Phys. Lett. {\bf A16}, 1327 (2001) [hep-th/0105249].
 
\bibitem{Med1} A. J. M. Medved, {\it Bad News on the Brane}, hep-th/0205251 ; 
{\it CFT on the brane with
                a Reissner-Nordstrom-de Sitter twist}, hep-th/0111182.
 
\bibitem{braneC} I. Savonije and E. Verlinde, Phys. Lett. {\bf B507}, 305 (2001) [hep-th/0102042];
                  Y. S. Myung, {\it Radially infalling brane and moving domain wall in the 
brane cosmology}, hep-th/0102184;
                 N. J. Kim, H. W. Lee, and Y. S. Myung, Phys. Lett. {\bf B504}, 323 (2001) [hep-th/0101091].
 
                 
\bibitem{Wald} R. M. Wald, {\it General Relativity} (University of Chicago Press, Chicago, 1984).
 
\bibitem{LorVio} C. Csaki, J. Erlich, and C. Grojean, Nucl. Phys. {\bf B604}, 312 (2001)[hep-th/0012143];
                 C. Csaki, {\it Asymmetrically Warped Spacetimes}, hep-th/0110269;
                 H. Stoica, {\it Comment on 4D Lorentz invariance violations in the brane-world}, hep-th/0112020.
 
\bibitem{BV2} C. Barcelo and M. Visser, {\it Twilight for the energy conditions?}, gr-qc/0205066.
 
\bibitem{BDL} P. Binetruy, C. Deffayet and D. Langlois, Nucl. Phys.
             {\bf B565}, 269 (2000)[hep-th/9905012]; P. Binetruy, C. Deffayet, U. Ellwanger and D. Langlois,
             Phys. Lett. {\bf B477}, 285 (2000)[hep-th/9910219].
 
\bibitem{STW} H. Stoica, S.-H. H. Tye and I. Wasserman, Phys. Lett. {\bf B482}, 205 (2000) [hep-th/0004126].
 
\bibitem{Kim} N.J. Kim, H.W. Lee, Y.S. Myung and G. Kang, Phys. Rev. {\bf D64}, 
064022 (2001) [hep-th/0104159].
 
\bibitem{GP} J. P. Gregory and A. Padilla, Class. Quant. Grav. {\bf 19}, 4071 (2002) [hep-th/0204218].
 
\bibitem{RS}L. Randall and R. Sundrum, Phys. Rev. Lett. {\bf 83}, 4690 (1999) [hep-th/9906064].
 
\bibitem{MAL} J. Maldacena, Adv. Theor. Math. Phys. {\bf 2}, 231 (1998) 
                [hep-th/9711200].
\bibitem{GKPW} S. S.Gubser, I. Klebanov, A. Polyakov, Phys. Lett.{\bf B428}, 105 (1998) [hep-th/9802109]; 
               E. Witten, Adv. Theor. Math. Phys. {\bf 2}, 253(1998) [hep-th/9802150].
 
\bibitem{Witten} E. Witten, Adv. Theor. Math. Phys. {\bf 2}, 505 (1998) [hep-th/9803131].
 
\bibitem{Cai} R. G. Cai, Phys. Rev. {\bf D63}, 124018 (2001)  [hep-th/0102113].
 
\bibitem{Noji} S. Nojiri and S. Odintsov, Phys. Lett. {\bf B494}, 135 (2000) [hep-th/0008160];
               Class. Quant. Grav. {\bf 18}, 5227 (2001) [hep-th/0103078].
 
\bibitem{Ver}E. Verlinde, {\it On the Holographic Principle in a Radiation Dominated Universe}, hep-th/0008140.
 
\bibitem{ISR} W. Israel, Nuovo Cim. {\bf B44}, 1 (1966); {\it ibid}. {\bf B48}, 463 (1967) .
 
\bibitem{Myung2} Y. S. Myung, Phys. Lett. {\bf B 531}, 1(2002) [hep-th/0112140].
 
\bibitem{BFKMP} T. Banks, W. Fischler, A. Kashani-Poor, R. McNees, and S. Rabin,
                 {\it Entropy of the Stiffest Stars}, hep-th/0206096.
 
\end{thebibliography}
\end{document}